# Seeking the Equation of State of Non-Compact Lattice QED

M. Göckeler[a,b], R. Horsley[c], V. Linke[d], P.E.L. Rakow[a],
G. Schierholz[e,a] and H. Stüben[f]

[a] Höchstleistungsrechenzentrum HLRZ,
c/o Forschungszentrum Jülich, D-52425 Jülich, Germany

[b] Institut für Theoretische Physik,
RWTH Aachen, D-52056 Aachen, Germany

[c] Institut für Physik, Humboldt-Universität zu Berlin,
Invalidenstraße 110, D-10115 Berlin, Germany

[d] Institut für Theoretische Physik, Freie Universität Berlin,
Arnimallee 14, D-14195 Berlin, Germany

[e] Deutsches Elektronen-Synchrotron DESY,
Notkestraße 85, D-22603 Hamburg, Germany

[f] Konrad-Zuse-Zentrum für Informationstechnik Berlin,
Heilbronner Straße 10, D-10711 Berlin, Germany

**Abstract**

We perform a high statistics calculation of the equation of state for non-compact QED on large lattices. The calculation extends to fermionic correlation lengths of $\approx 8$, and it is combined with a finite size scaling analysis of the lattice data.

# 1 Introduction

Non-compact lattice QED undergoes a second order chiral phase transition at strong coupling [1]. This allows one to take the cut-off to infinity, which is prerequisite to a non-perturbative definition of a continuum theory. A wide-spread belief is that QED, like other non asymptotically-free theories, is trivial in the sense that all renormalized couplings vanish as the cut-off is taken to infinity. It would, of course, be much more interesting if the theory had a non-trivial continuum limit.

In a series of papers [2, 3, 4, 5, 6] we have investigated the ultra-violet behaviour of four-flavour non-compact QED using staggered fermions. Among other things, we found that the data are consistent with mean field critical exponents plus logarithmic corrections, and that the renormalized charge vanishes in the continuum limit. Furthermore, the effective Yukawa couplings could be shown to follow roughly the behaviour of the renormalized charge, which suggests that they too vanish at the critical point.

On the other hand, the Illinois group has mainly focused on the equation of state. They first reported [7] Miransky scaling [8] and argued that the phase transition is driven by monopole condensation, leading to electron confinement in the chirally broken phase [9]. Later on this transition was interpreted as four-dimensional percolation [10] with power-law critical exponents, the major support for this idea being the apparent coincidence of chiral phase transition and monopole percolation thresholds. But "the truth cannot be so simple", as Hands and Kogut concluded [11] correctly [12].

Another approach was suggested by the Zaragoza group, who use a mean field guided algorithm to simulate the effect of dynamical fermions [13, 14, 15]. In this algorithm the chiral limit is taken before the infinite volume limit. It is known from examples of explicitly solvable models that this can lead to wrong results [16]. We expect this procedure to be particularly dangerous in the symmetric phase of the theory.

In spite of these efforts the subject has remained controversial. The various groups disagree in the exact position of the critical point and the critical exponents of the chiral phase transition. In the present paper we shall return to the determination of the equation of state with improved statistics, simulating on larger lattices and closer to the critical point, and hopefully answer some of the open questions.

The paper is organized as follows. In section 2 we give the details of the lattice calculation. In section 3 we derive a finite size formula for the chiral condensate and the ratio of scalar and pseudoscalar susceptibilities. We use that formula to extrapolate the lattice data to the infinite volume. In section 4 we investigate possible forms of the equation of state, and in section 5 we determine the parameters by fitting to the data. Alternatively, the critical coupling and exponents can be determined from the susceptibility ratio, which is discussed in section 6. In section 7 we derive a relation between the equation of state and the average plaquette. This provides a further test of our results. Finally, in section 8 we end with some concluding remarks.

## 2  Lattice Calculation

The non-compact gauge field action is given by

$$S_G = \frac{\beta}{2} \sum_{x,\mu<\nu} F_{\mu\nu}^2(x) \tag{1}$$

with

$$F_{\mu\nu}(x) = \Delta_\mu A_\nu(x) - \Delta_\nu A_\mu(x), \tag{2}$$

where $\Delta_\mu$ is the forward lattice derivative, $\beta = 1/e^2$, and $e$ is the bare charge. In Eqs. (1) and (2) and in the following the lattice constant has been set equal to 1 for convenience, so that all dimensionful quantities are to be understood in units of the (inverse) lattice spacing. The gauge fields take values on the real line. As long as one only considers gauge invariant quantities, the functional integral can always be made well-behaved, in spite of the unbounded range of integration.

Since chiral symmetry plays a major role in this work, a natural choice for the fermionic variables are staggered fermions. The corresponding action is given by

$$S_F = \sum_{x,y} \bar{\chi}(x) M_{xy} \chi(y), \tag{3}$$

$$\begin{aligned} M_{xy} &= m\delta_{xy} + \frac{1}{2}\sum_\mu (-1)^{x_1+\ldots+x_{\mu-1}} [e^{iA_\mu(x)}\delta_{y\,x+\hat{\mu}} - e^{-iA_\mu(y)}\delta_{y\,x-\hat{\mu}}] \\ &\equiv m\delta_{xy} + iD_{xy}, \end{aligned} \tag{4}$$

where $m$ is the bare mass. In the naive continuum limit this action describes four Dirac fermions (flavours) minimally coupled to a $U(1)$ gauge field. For finite lattice spacing it has a chiral $U(1) \times U(1)$ symmetry at $m = 0$. The physically interesting region is near the phase transition at $\beta = \beta_c$, where this chiral symmetry is spontaneously broken.

The calculations in this paper are based on the action $S = S_G + S_F$, where we have used periodic boundary conditions for the gauge fields and periodic (anti-periodic) spatial (temporal) boundary conditions for the fermions. The extent of the lattice will be denoted by $L$, so that the four-dimensional volume is given by $V = L^4$.

We have performed simulations on $4^4$, $6^4$, $8^4$, $10^4$, $12^4$, $16^4$ and $20^4$ lattices at $\beta$ values ranging between 0.16 and 0.22 and at masses between 0.005 and 0.16. The actual values can be read off from Tables 1 – 3. We have used the hybrid Monte Carlo algorithm [17] for updating the gauge field configurations. Some details of the performance of the algorithm for QED can be found in Ref. [2].

We have in general accumulated 1000 trajectories for each pair of parameters $(\beta, m)$. At the smallest bare mass, $m = 0.005$, the statistics is in some cases only 500 trajectories, while for the $20^4$ lattice it is only 100 trajectories. On the $6^4$ lattice we generated 5000 trajectories for each $(\beta, m)$ and on the $4^4$ lattice 10000 trajectories for each $(\beta, m)$. The trajectory length was chosen to be 0.7–1.0, and the molecular dynamics step sizes were adjusted so that acceptance rates of 70–80 % were obtained. The stopping criterion for the conjugate gradient inverter was taken to be $r^2 < 10^{-10}$ for $m > 0.005$ and $r^2 < 10^{-14}$ for $m = 0.005$. Compared to our previous investigations this represents a considerable increase in statistics.

We have computed the average plaquette $P$, the chiral condensate $\sigma$ and the logarithmic derivative $R$ of the chiral condensate,

$$R \equiv \left.\frac{\partial \ln \sigma}{\partial \ln m}\right|_\beta = \left.\frac{m}{\sigma}\frac{\partial \sigma}{\partial m}\right|_\beta . \tag{5}$$

From rigorous Ward identities we know that $R$ can be independently computed as the ratio of zero momentum meson propagators (susceptibilities):

$$\left.\frac{\partial \sigma}{\partial m}\right|_\beta = C_\sigma(p=0),$$
$$\frac{\sigma}{m} = C_\pi(p=0), \tag{6}$$

so that

$$R = \frac{C_\sigma(p=0)}{C_\pi(p=0)}. \tag{7}$$

Naturally, this is only an exact identity if the full propagators are used:

$$C_\sigma(p=0) \equiv \sum_x \left(\langle \bar{\chi}(0)\chi(0)\bar{\chi}(x)\chi(x)\rangle - \langle \bar{\chi}(0)\chi(0)\rangle\langle \bar{\chi}(x)\chi(x)\rangle\right), \tag{8}$$

$$C_\pi(p=0) \equiv \sum_x (-1)^{x_1+x_2+x_3+x_4}\left(\langle \bar{\chi}(0)\chi(0)\bar{\chi}(x)\chi(x)\rangle\right.$$
$$\left. - \langle \bar{\chi}(0)\chi(0)\rangle\langle \bar{\chi}(x)\chi(x)\rangle\right)$$
$$= \sum_x (-1)^{x_1+x_2+x_3+x_4}\langle \bar{\chi}(0)\chi(0)\bar{\chi}(x)\chi(x)\rangle . \tag{9}$$

It can be shown that in the pion propagator at momentum zero only the fermion-line-connected part of $\langle \bar{\chi}(0)\chi(0)\bar{\chi}(x)\chi(x)\rangle$ contributes. This is not the case for $C_\sigma$, where the fermion-line-disconnected contribution is important. We found that in the broken phase it can change $C_\sigma$ by $\approx 50\,\%$.

We computed the fermionic observables, including the fermion-line-disconnected part of $C_\sigma$, by using stochastic estimators (see e.g. [18]). The computation is based on expressing Eq. (8) in terms of the fermion matrix

$M$ defined in Eq. (4),

$$C_\sigma(p=0) = \frac{1}{V} \sum_{x,y} \left( \langle \bar{\chi}(x)\chi(x)\bar{\chi}(y)\chi(y) \rangle - \langle \bar{\chi}(x)\chi(x) \rangle \langle \bar{\chi}(y)\chi(y) \rangle \right)$$
$$= \frac{1}{V} \left( \langle \text{Tr}\, M^{-1} \text{Tr}\, M^{-1} \rangle - \langle \text{Tr}(M^{-1}M^{-1}) \rangle - \langle \text{Tr}\, M^{-1} \rangle^2 \right). \quad (10)$$

We obtained the traces of these matrices by averaging over vectors of random numbers $\eta$,

$$\text{Tr}\, A = \langle \eta^\dagger A \eta \rangle_\eta. \quad (11)$$

The random variables have to be chosen such that $\langle \eta \rangle_\eta = 0$ and $\langle \eta_x^* \eta_y \rangle_\eta = \delta_{xy}$. In our calculations we used twenty random vectors for each configuration.

The results for the average plaquette $P$, the chiral condensate $\sigma$, and the ratio $R$ are shown in Tables 1, 2 and 3, respectively.

## 3  Finite Size Analysis

Important for analyzing the data is an understanding of the finite size effects. Although in general they are small, results in the symmetric phase and at smaller bare masses can suffer from finite size effects, $\sigma$ being under-estimated on small lattices.

According to finite size-scaling theory [19, 20] we expect the general form of the chiral condensate $\sigma$ and correlation length $\xi$ to be given by scaling relations of the form

$$\sigma = L^{-\dot{\beta}/\nu} \tilde{\sigma}\left(\tau L^{1/\nu}, m L^{\delta\dot{\beta}/\nu}\right) \text{ and } \xi = L\, \tilde{\xi}\left(\tau L^{1/\nu}, m L^{\delta\dot{\beta}/\nu}\right), \quad (12)$$

where $\tau = (\beta - \beta_c)/\beta_c$ and $\dot{\beta}$, $\delta$ and $\nu$ are the critical exponents. We have used the symbol $\dot{\beta}$ for the critical exponent $\beta$ to distinguish it from the inverse coupling $\beta$. The identity $\gamma = \dot{\beta}(\delta - 1)$ has been used to eliminate the critical exponent $\gamma$ in favour of $\delta$ in the above expressions. These scaling relations can be rearranged to give an implicit equation which the chiral condensate must satisfy,

$$f\left(L/\xi, \tau/\sigma^{1/\dot{\beta}}, m/\sigma^\delta\right) = f\left(m_R L, \tau/\sigma^{1/\dot{\beta}}, m/\sigma^\delta\right) = 0, \quad (13)$$

where $m_R$ is the renormalized mass.

Unfortunately, this form is far too general for our purposes. Therefore we need to make a model calculation to see what form $f$ has. Because finite size effects are largest at large $\beta$, we have investigated them in the $\beta = \infty$ limit, where they are exactly calculable.

At infinite $\beta$ all plaquettes are forced to have minimum action, so the only gauge fields we have to consider are those that are gauge-equivalent to a constant $A$ field. We need to know the fermion determinant in such a background field. The fermion determinant $F$ for $N_f$ flavours of staggered fermions is

$$F \equiv \det (m + \mathrm{i} D)^{(N_f/4)} = \det \left(m^2 + D^2\right)^{(N_f/8)}. \tag{14}$$

It is more convenient to work with the second form, because the eigenvectors of $D^2$ are simpler than those of $D$. The eigenvectors of $D^2$ in a constant $A$ field are simply $\exp(-\mathrm{i} k \cdot x)$, where the momentum vector $k$ is consistent with the boundary conditions. The eigenvectors do not depend on the value of $A$, however the corresponding eigenvalues do depend on the background field. They have the value $\sum_\mu \sin^2(k_\mu - A_\mu)$. Thus

$$F(A) = \prod_k (K(k-A))^{(N_f/8)} = \exp\left(\frac{N_f}{8} \sum_k \ln K(k-A)\right), \tag{15}$$

where

$$K(p) \equiv m^2 + \sum_\mu \sin^2(p_\mu). \tag{16}$$

We can easily calculate the chiral condensate from Eq. (15):

$$\sigma = \frac{4}{V N_f} \frac{\partial}{\partial m} \int d^4 A\, F(A) = \frac{\int d^4 A\, F(A)\, s(A)}{\int d^4 A\, F(A)} \tag{17}$$

with

$$s(A) = \sum_k \frac{m}{K(k-A)}. \tag{18}$$

Equation (17) is exact, and we have evaluated it on the computer, but it is more useful to look at the limit $m \ll 1$, $mL \gg 1$, where we can simplify this expression. We find the large $L$ limit of the $k$ sums in the standard way, using identities of the form

$$\sum_{n=1}^{L} g\left(\frac{2\pi n}{L} - \theta\right) = L \sum_{j=-\infty}^{\infty} \exp(\mathrm{i} j L \theta) \int_0^{2\pi} \frac{dp}{2\pi}\, g(p) \exp(\mathrm{i} p j L), \tag{19}$$

valid for a periodic function $g(k)$ with period $2\pi$. Applying the four-dimensional analogue of Eq. (19) to the sums of interest gives

$$s(A) = \sum_j \exp(\mathrm{i} L j \cdot A) \int \frac{d^4 p}{(2\pi)^4} \exp(\mathrm{i} L j \cdot p) \frac{m}{K(p)}. \tag{20}$$

When the four-vector $j$ is non-zero, the $p$ integral is exponentially small. The leading behaviour of these integrals can be found in several ways, for example by the saddle point approximation, or by considering the asymptotic behaviour of the Bessel functions:

$$s(A) \approx \int \frac{d^4 p}{(2\pi)^4} \frac{m}{K(p)} + \sum_{j \neq 0} \exp(\mathrm{i} L j \cdot A) \left(\frac{2m}{\pi L |j|}\right)^{\frac{3}{2}} \exp(-m L |j|). \tag{21}$$

Similarly, for the sum of logarithms appearing in Eq. (15) we find

$$\frac{1}{V}\sum_k \ln K(k-A) = \sum_j \exp(\mathrm{i}Lj \cdot A) \int \frac{d^4p}{(2\pi)^4} \exp(\mathrm{i}Lj \cdot p) \ln K(p)$$
$$\approx \int \frac{d^4p}{(2\pi)^4} \ln K(p) - \sum_{j\neq 0} \exp(\mathrm{i}Lj \cdot A) \frac{2}{L|j|} \left(\frac{2m}{\pi L|j|}\right)^{\frac{3}{2}} \exp(-mL|j|). \quad (22)$$

Substituting Eqs. (21), (22) into Eq. (17) and collecting terms independent of $A$ (terms with cyclic dependence on $A$ give zero when $A$ is integrated over), we get

$$\sigma = \int \frac{d^4p}{(2\pi)^4} \frac{m}{K(p)} - \frac{16}{\pi^3} N_f m^3 \exp(-2mL) + \cdots$$
$$= \sigma_\infty - \frac{16}{\pi^3} N_f m^3 \exp(-2mL) + \cdots. \quad (23)$$

Features of Eq. (23) to note are that no terms of order $\exp(-mL)$ survive the integration over the background fields $A$, and that in the coefficient of $\exp(-2mL)$ all powers of $L$ cancel, leaving a coefficient $\propto L^0$. This formula has been derived at $\beta = \infty$, where bare and renormalized fermion masses are the same. We want to use it to suggest a finite size formula which can be used at finite $\beta$ too, so we have to consider whether to interpret $m$ as the bare or renormalized mass. On physical grounds it is clear that the mass appearing in the exponential function should be the renormalized mass $m_R$. We have therefore used the observation [3] that the lowest-order result $\sigma \approx 0.62 m_R$ works well, both at $\beta = \infty$ and in the critical region, to rewrite the formula in terms of $\sigma$, a quantity easier to calculate than $m_R$,

$$\sigma - \sigma_\infty \propto N_f\, m_R^3 \exp(-2m_R L) + \cdots$$
$$\propto N_f\, \sigma^3 \exp(-3.23\,\sigma L) + \cdots. \quad (24)$$

This formula was applied in [4, 5] successfully. Obviously, a formula derived at $\beta = \infty$ must be tested before being applied at other $\beta$ values. If we plot $\sigma$ against the right hand side of Eq. (24), we obtain straight lines [5]. Another encouraging observation is that finite size effects are tiny in the quenched case and grow as the number of flavours, $N_f$, increases [10], as expected from Eq. (24).

In order to reconcile this calculation with the *ansatz* (13), we write

$$\sigma_\infty = \sigma\left(1 + a(\sigma, m, L)\right), \quad (25)$$

where

$$a(\sigma, m, L) = A\left(\frac{\sigma^\delta}{m}\right)^q \exp(-3.23\,\sigma L). \quad (26)$$

Differentiating this finite size formula for $\sigma$ gives the following result for $R$:

$$R_\infty = \frac{R\left[1 + a(\sigma, m, L)(1 + \delta q - 3.23\,\sigma L)\right] - q a(\sigma, m, L)}{1 + a(\sigma, m, L)}. \quad (27)$$

The prefactor is now linear in $L$, which indicates that the finite size corrections to $R$ will be larger than those for $\sigma$.

In order to obtain values for $\sigma_\infty$ and $R_\infty$, we made combined overall fits to the available data (with the restrictions $L \geq 8$, and $3.23\,\sigma L > 3$ which corresponds to $m_R L > 1.5$). In total 138 values for $\sigma$ and 49 values for $R$ were used in the fit. We only have three free parameters, $A$, $\delta$ and $q$, to describe all these data. The results of the fit are $A = 29.7(35)$, $q = 0.754(23)$ and $\delta q = 2.465(56)$. Hence we obtain the critical exponent $\delta = 3.27(12)$. The $\chi^2/\text{dof}$ is 2.29.

Plots of this fit are shown in Figs. 1 and 2. The data are consistent with the formula. We see that our restriction on using $\sigma$ data was quite conservative. Our *ansatz* describes the $\sigma$ data down to $2m_R L \approx 3.23\,\sigma L = 2$, and is also valid for many data from the smaller lattices.

As a check of our assumption $\sigma \approx 0.62 m_R$, we experimented with introducing a fourth parameter $\varepsilon$ in the exponential term $\exp(-\varepsilon\,3.23\,\sigma L)$. It came out to be 1 within 1 % error, showing that such a modification of the exponential term is not needed.

The results of the extrapolation to infinite volume are given in the last column of Tables 2 and 3. In most cases $\sigma_\infty$ differs very little from the value on the largest lattice we used.

## 4  Equation of State

In this section we will attempt to numerically determine the equation of state

$$m = f(\sigma, \beta). \tag{28}$$

From now on, we take $\sigma$ to be the value extrapolated to the infinite volume, as given by the last column in Table 2. Noting that for $m$ non-zero $f$ is analytic in $\beta$, we expand it around $\beta = \beta_c$. This gives

$$m = f_0(\sigma) + (\beta - \beta_c)f_1(\sigma) + O((\beta - \beta_c)^2). \tag{29}$$

The function $f_0(\sigma)$ must vanish faster than $f_1(\sigma)$ for $\sigma \to 0$. To test whether it is reasonable to truncate the series after $f_1$, we look at a contour plot obtained by interpolating the infinite volume values of $\sigma$. We interpolate linearly in the variable $\beta$ and logarithmically in $m$. The results are shown in Fig. 3. As we see very little curvature, we conclude that higher terms in the expansion are negligible.

We fit a line to each contour independently, and find $f_0$ and $f_1$ from the intercepts and gradients. This allows us to investigate the functions $f_0$ and $f_1$ without having to make any assumptions about the form of their $\sigma$ dependence. To investigate deviations from mean field behaviour

$$\begin{align} f_0(\sigma) &\propto \sigma^3, \\ f_1(\sigma) &\propto \sigma, \end{align} \tag{30}$$

we plot in Fig. 4 $f_1(\sigma)/\sigma$ versus $\sigma$. We see that $f_1$ grows slightly faster than $\sigma$. Since this is a log-log plot, the curvature in the data indicates a non-power behaviour of $f_1$.

Knowing $f_1$, we can estimate $\beta_c$. We rewrite Eq. (29) as

$$\beta - \frac{m}{f_1(\sigma)} = \beta_c - \frac{f_0(\sigma)}{f_1(\sigma)}. \tag{31}$$

The ratio $f_0/f_1$ vanishes as $\sigma \to 0$, in the mean field case quadratically. This suggests plotting the l.h.s. of Eq. (31) against $\sigma^2$ to get $\beta_c$ from the intercept. From Fig. 5 we read off $\beta_c \approx 0.19$. Deviations from $\sigma^2$ behaviour are small.

Finally, we attempt to determine $f_0(\sigma)$ by rearranging the equation of state (29) to read

$$f_0(\sigma) = m - (\beta - \beta_c)f_1(\sigma). \tag{32}$$

Knowing $f_1$ and $\beta_c$ gives in principle $f_0$. Of course, the result is somewhat uncertain at the smallest $\sigma$ values due to the sensitivity on $\beta_c$. In Fig. 6 we show, motivated from mean field exponents (30), $f_0(\sigma)/\sigma^3$ against $\sigma$ for $\beta_c = 0.19040(9)$ (the value to be found in the following section).

## 5 Critical Behaviour

In this section we study specfic *ansätze* for the functions $f_0(\sigma)$ and $f_1(\sigma)$ introduced in Eq. (29). We start with the mean field equation of state

$$m = A_0 \sigma^3 + A_1(\beta - \beta_c)\sigma. \tag{33}$$

As we have already seen in the previous section, there are corrections to mean field behaviour. Nevertheless, it is instructive to look at a Fisher plot [21], because it gives another perspective. This is shown in Fig. 7. From Eq. (33) we expect to see straight parallel lines if we plot $\sigma^2$ over $m/\sigma$. These lines are labeled by $\beta$, and the line corresponding to $\beta_c$ ends at the origin. In Fig. 7 we see just such behaviour. We can read off $0.19 < \beta_c < 0.195$. The curvature of these lines is slight, showing that the deviations from mean field behaviour are not large.

There are two *ansätze* for modifications to Eq. (33) discussed in the literature. Our approach has been the introduction of logarithmic corrections [2]:

$$m = A_0 \frac{\sigma^3}{\ln^{p_0}(1/\sigma)} + A_1(\beta - \beta_c)\frac{\sigma}{\ln^{p_1}(1/\sigma)}. \tag{34}$$

This equation has five parameters, namely $\beta_c, A_0, p_0, A_1$ and $p_1$. An alternative *ansatz* used in [10] is

$$m = A_0 \sigma^\delta + A_1(\beta - \beta_c)\sigma^b. \tag{35}$$

It has also five parameters, namely $\beta_c, A_0, \delta, A_1$ and $b$. This *ansatz* means that one is looking for non-mean field critical exponents $\delta$ and $\dot\beta$ (recall that $b = \delta - 1/\dot\beta$, where again we have denoted the critical exponent $\beta$ by $\dot\beta$ to distinguish it from the inverse coupling $\beta$).

We fit both equations to the data, as we did already in [5]. In addition to the increase in statistics, there is improvement in two directions. Firstly, we have now extrapolated our data to infinite volume. Secondly, we have made simultaneous fits to $\sigma$ and $R$ data, using all the independent information that we have. According to the definition (5), the fit formula for $R$ can be obtained by differentiating Eqs. (34) and (35). We fitted all infinite volume data in the parameter range $m \leq 0.05$ and $0.16 \leq \beta \leq 0.22$ (see Tables 2 and 3). The total number of data used is 77 for $\sigma$ and 42 for $R$. The results are:

| Fit | $\beta_c$ | $A_0$ | $p_0$ | $A_1$ | $p_1$ | $\chi^2/\text{dof}$ |
|---|---|---|---|---|---|---|
| 1 | 0.19040(9) | 1.798(5) | 0.324(15) | 6.76(3) | 0.485(7) | 7.63 |
| 2 | 0.18748(4) | 1.828(6) | 1 (fixed) | 7.46(2) | 0.686(5) | 20.55 |
| Fit | $\beta_c$ | $A_0$ | $\delta$ | $A_1$ | $b$ | $\chi^2/\text{dof}$ |
| 3 | 0.19039(11) | 2.138(29) | 3.206(14) | 8.154(55) | 1.255(4) | 9.72 |
| 4 | 0.19617(8) | 1.203(8) | 2.596(7) | 5.300(8) | 1 (fixed) | 47.86 |

The errors given in this table are directly taken from MINUIT. In Fig. 8 we show $\sigma$ data together with fit 1. Figure 9 shows this fit compared with $R$ data. Several comments are in order:

- Fit 2 is included to compare with our old result, where we fixed $p_0 \equiv 1$ and found $\beta_c = 0.187(1)$, $p_1 = 0.61(2)$ [3]. Our new data prefer a different $p_0$: leaving $p_0$ as an additional free parameter improves the $\chi^2/\text{dof}$ considerably.

- Fixing $b \equiv 1$, as we did in fit 4, was proposed in [10]. Compared with fit 3, the $\chi^2/\text{dof}$ increases by a factor of nearly 5.

- Fits 1 and 3 give consistent results, especially for $\beta_c$. This is to be expected because Eqs. (34) and (35) are numerically similar. From Eq. (34) one can derive effective values for $\delta$ and $b$, $\delta_{\text{eff}} = 3 + p_0/\ln(1/\sigma)$, $b_{\text{eff}} = 1 + p_1/\ln(1/\sigma)$. The positive values of $p_0$ and $p_1$ correspond to $\delta > 3$ and $b > 1$.

- Finally, we can compare the fit with our findings of the previous section. In Figs. 4, 5 and 6 we plotted the results from fit 1 (solid lines) and fit 3 (dotted lines). Data and fits differ mainly for large values of $\sigma$. In Figs. 4 and 5 fit 1 describes the data a little better than fit 3. From Fig. 4 we also see that $f_1/\sigma$ is not constant. This would be the case if $b = 1$.

We see that $\beta_c$ is insensitive to the specific *ansatz* used for the equation of state. The data do not support the assumption $b = 1$.

# 6  Susceptibility Ratio

An alternative approach to the determination of $\beta_c$ and $\delta$ was proposed in [10]. It is based on the susceptibility ratio $R$ defined in Eq. (5).

The critical behaviour of the equation of state can be read off from the form of $R$ at small $m$ values. In the broken phase $R$ must vanish when $m \to 0$. If $\sigma$ is proportional to $m^p$ for small $m$, as expected at the phase transition and (with a different power) in the symmetric phase, $R$ will go to $p$ as $m \to 0$. At $\beta_c$ the power $p$ is $1/\delta$, so we can determine $\delta$ from the intercept of the $R$ curve at the critical coupling. In the symmetric phase $p$ is $1/(\delta - 1/\dot\beta)$. In both cases logarithmic corrections, as in Eq. (34), can mean that the asymptotic values are only reached at extremely low $m$. Since at large $m$ the fermion determinant and the partition function are $\sim m^V$, we know that $\sigma \to 1/m$ at large $m$. Therefore, at very large $m$, $R$ goes to $-1$ for all $\beta$.

In Fig. 9 we plot the $R$ values extrapolated to infinite volume according to Eq. (27). The separatrix between curves showing the behaviour expected in the broken and the symmetric phase, respectively, corresponds to a coupling which lies slightly above 0.19. This is consistent with the results of sections 4 and 5. The exponent $\delta$ is in the neighbourhood of 3.

In Fig. 9 we also compare the $R$ values with the equation of state derived in section 5. The data are compatible with the fit, though naturally the $\sigma$ data with their smaller statistical errors and finite size effects dominate the determination of the critical parameters.

In [10] the mass ratio $m_\pi^2/m_\sigma^2$ was used as an approximation to the susceptibility ratio $C_\sigma/C_\pi$. The approximations involved in this substitution are discussed in [4]. Comparing the $R$ values in Table 3 with the mass ratios in [10], we see that this is not a valid approximation.

# 7  Maxwell Relation

Another quantity which can be computed with high precision is the average plaquette $P$,

$$P \equiv \frac{1}{6V} \left\langle \sum_{x,\mu<\nu} F_{\mu\nu}^2(x) \right\rangle . \qquad (36)$$

The plaquette values can be related to the equation of state for the chiral condensate by means of a Maxwell relation, which we now derive.

We know that both the chiral condensate and the average plaquette can be found from the partial derivatives of the partition function $Z$:

$$\left. \frac{1}{V} \frac{\partial}{\partial m} \ln Z \right|_\beta = \frac{N_f}{4} \sigma \qquad (37)$$

and
$$\left.\frac{1}{V}\frac{\partial}{\partial\beta}\ln Z\right|_m = -6P\,. \tag{38}$$

This allows us to express the second derivative $\frac{\partial}{\partial m}\frac{\partial}{\partial\beta}\ln Z$ in two different ways:

$$\frac{1}{V}\frac{\partial}{\partial m}\frac{\partial}{\partial\beta}\ln Z = \left.\frac{N_f}{4}\frac{\partial\sigma}{\partial\beta}\right|_m = \left.-6\frac{\partial P}{\partial m}\right|_\sigma\,, \tag{39}$$

which leads to

$$\left.\frac{\partial P}{\partial m}\right|_\beta = -\left.\frac{N_f}{24}\frac{\partial\sigma}{\partial\beta}\right|_m = \left.\frac{N_f}{24}\frac{\partial m}{\partial\beta}\right|_\sigma \left.\frac{\partial\sigma}{\partial m}\right|_\beta\,. \tag{40}$$

These thermodynamic relations hold for any lattice size and for all values of $\beta$ and $m$. The second form of the identity is the most useful, as our equations of state give us $m$ as a function of $\sigma$ and $\beta$. For this reason we will express $P$ as a function of $\sigma$ and $\beta$, too.

The partial differential equation, Eq. (40), can be solved by

$$P(\sigma,\beta) = \frac{N_f}{24}\int_0^\sigma d\sigma'\frac{\partial f(\sigma',\beta)}{\partial\beta} + C(\beta) \equiv I(\sigma,\beta) + C(\beta)\,, \tag{41}$$

where $m = f(\sigma,\beta)$ is the equation of state and $C(\beta)$ is a constant of integration, which can depend only on $\beta$. If terms of $O((\beta - \beta_c)^2)$ in Eq. (29) are neglected (as in the *ansätze* (34) and (35)), the integral $I$ is independent of $\beta$ and depends only on $\sigma$. Applying Eq. (41) to the equation of state (34) leads to

$$P(\sigma,\beta) - \frac{N_f}{24}A_1 2^{p_1-1}\Gamma\left(1-p_1, 2\ln(1/\sigma)\right) = C(\beta)\,, \tag{42}$$

where $\Gamma$ is the incomplete $\Gamma$ function [22]. The corresponding result from the power-law *ansatz* (35) is

$$P(\sigma,\beta) - \frac{N_f}{24}\frac{A_1}{b+1}\sigma^{b+1} = C(\beta)\,. \tag{43}$$

Using the plaquette values reported in Table 1, we have plotted in Fig. 10 $\beta(P(\sigma,\beta) - I(\sigma))$, i.e. the l.h.s. of Eqs. (42) and (43) multiplied by $\beta$. If the equation of state is accurate, we should find that $\beta(P - I)$ depends only on $\beta$, so the values calculated at different $m$ should all lie on a single curve. We find that the logarithmic equation of state (fit 1) and the power-law equation of state with all exponents free (fit 3) satisfy this test. However, in the power-law fit with $b \equiv 1$ (fit 4), shown in Fig. 11, we see deviations depending systematically on $m$.

Azcoiti et al. have used their technique to compute the plaquette values at $m = 0$ on an $8^4$ lattice [13, 14]. These data have to be regarded with some caution, because of the small lattice size, and because (as already noted) it would be preferable to take the large volume limit before the $m \to 0$ limit.

Nevertheless, we have compared the data reported in [13] with the expected infinite volume value given by the Maxwell relation (42). The results are shown in Fig. 12. The agreement is fair. The $8^4$ data differ by less than 1 % from our infinite volume expectation. It would be most interesting to know whether the values found on larger lattices move in the expected direction. (Simulations have recently been carried out on larger lattices [23]. Unfortunately plaquette values are not reported.) In [13, 14] a critical coupling of $\beta_c = 0.208(4)$ is extracted from these calculations. In view of Fig. 12, we think the data are just as compatible with $\beta_c = 0.19$.

## 8 Conclusions

We have presented a determination of the equation of state, including the critical exponents and coupling, of four-flavour non-compact QED. This work extends previous investigations in several respects. Most important was the finite size analysis of the lattice data. The finite size formula, which we have derived for the chiral condensate $\sigma$ and the susceptibility ratio $R$, was found to be in good agreement with the numerical results. This allowed us to extrapolate the numbers to the infinite volume. A further addition was that we fitted the $\sigma$ and $R$ data simultaneously.

With the finite size formula at hand, we can compare our data with the results of other groups taken on different-sized lattices. We find good agreement with the results of the Illinois group. These authors use a hybrid Monte Carlo algorithm like ours. The Zaragoza group has only published data on $8^4$ lattices. Their $\sigma$ values are consistent with the prediction of the finite size formula, albeit a precise comparison is not possible because of the relatively large errors of the data. Their plaquette values at $m = 0$ on $8^4$ lattices differ by less than 1 % from our extrapolated infinite lattice values.

The magnitude of finite size effects can be read off from Tables 1, 2 and 3. We see that finite size effects in the chiral condensate can be neglected, compared to the statistical errors, only if $\sigma L \gtrsim 3$. A large portion of the data that went into a recent analysis of the Zaragoza group [23] does not satisfy this constraint. Since finite size effects are largest when $m$ is small, one must be particularly cautious about zero mass measurements.

We have taken special care in determining the critical coupling $\beta_c$. In the Fisher plot (Fig. 7), susceptibility ratio plot (Fig. 9) and in Fig. 5 we can see directly from the data, without making fits, that the critical coupling is $\beta_c \approx 0.19$. Both the logarithmic (fit 1) and power-law equation of state (fit 3) give $\beta_c = 0.1904(1)$.

The Maxwell relation provides an independent test of the equation of state. The logarithmic fit (fit 1), as well as the unconstrained power-law fit (fit 3), passed this test successfully.

For simplicity we have made use of the power-law scaling relation in our finite size analysis. Assuming logarithmically improved mean field behaviour would give the same result. For the critical exponent $\delta$ that enters the finite size scaling formula (25) we find the effective value $\delta = 3.27(12)$. This is in excellent agreement with the result of the power-law fit (fit 3) of $\delta = 3.206(14)$.

Our final conclusion is that the data are consistent with a logarithmically improved mean field equation of state, as one would expect for a trivial theory. However, a power-law equation of state can describe the data nearly as well. A more direct approach to the problem, though much more demanding, is to determine the renormalized couplings of the theory near and at the critical point [2, 6]. A refined calculation of the renormalized charge on larger lattices and closer to the transition point is in progress.

# Acknowledgement

Our new simulations were done on the Siemens/Fujitsu S400/40 at RRZN Hannover and on the CRAY T3D at Konrad-Zuse-Zentrum für Informationstechnik Berlin. We thank Konrad-Zuse-Zentrum for generously providing CPU time on these machines.

# References

[1] For reviews and further references see:
    G. Schierholz, Nucl. Phys. **B** (Proc. Suppl.) **20** (1991) 623;
    A. Kocić, Nucl. Phys. **B** (Proc. Suppl.) **34** (1994) 123.

[2] M. Göckeler, R. Horsley, E. Laermann, P. Rakow, G. Schierholz, R. Sommer and U.-J. Wiese, Nucl. Phys. **B334** (1990) 527.

[3] M. Göckeler, R. Horsley, P. Rakow, G. Schierholz and R. Sommer, Nucl. Phys. **B371** (1992) 713.

[4] M. Göckeler, R. Horsley, P. Rakow, G. Schierholz and H. Stüben, Nucl. Phys. B (Proc. Suppl.) **34** (1994) 527.

[5] M. Göckeler, R. Horsley, V. Linke, P. Rakow, G. Schierholz and H. Stüben, Nucl. Phys. B (Proc. Suppl.) **42** (1995) 660.

[6] M. Göckeler, R. Horsley, P. Rakow and G. Schierholz, Phys. Lett. **B353** (1995) 100.

[7] J. B. Kogut, E. Dagotto and A. Kocić, Phys. Rev. Lett. **60** (1988) 772; Nucl. Phys. **B317** (1989) 253; **B317** (1989) 271.

[8] V. A. Miransky, Nuovo Cim. **A90** (1985) 149; Sov. Phys. JETP **61** (1985) 905.

[9] S. J. Hands, J. B. Kogut, R. Renken, A. Kocić, D. K. Sinclair and K. C. Wang, Phys. Lett. **B261** (1991) 294.

[10] A. Kocić, J. B. Kogut and K. C. Wang, Nucl. Phys. **B398** (1993) 405.

[11] S. Hands and J. B. Kogut, Nucl. Phys. **B462** (1996) 291.

[12] M. Göckeler, R. Horsley, P. Rakow and G. Schierholz, Phys. Rev. **D53** (1996) 1508.

[13] V. Azcoiti, G. Di Carlo and A. F. Grillo, Int. J. Mod. Phys. **A8** (1993) 4235.

[14] V. Azcoiti, G. Di Carlo and A. F. Grillo, Nucl. Phys. **B** (Proc. Suppl.) **30** (1993) 745.

[15] V. Azcoiti, G. Di Carlo, A. Galante, A. F. Grillo, V. Laliena and C. E. Piedrafita, Phys. Lett. **B353** (1995) 279.

[16] J. E. Hetrick, Y. Hosotani and S. Iso, Tucson preprint AZPH-TH/95-25 (1995) (`hep-th/9510090`).

[17] S. Duane, A. D. Kennedy, B. J. Pendleton and D. Roweth, Phys. Lett. **B195** (1987) 216.

[18] K. Bitar, A. D. Kennedy, R. Horsley, S. Meyer and P. Rossi, Nucl. Phys. **B313** (1989) 348.

[19] V. Privman (ed.), *Finite Size Scaling and Numerical Simulation of Statistical Systems* (World Scientific, Singapore, 1990).

[20] K. Binder, Annu. Rev. Phys. Chem. **43** (1992) 33.

[21] J. S. Kouvel and M. E. Fisher, Phys. Rev. **136** (1964) A 1626.

[22] M. Abramowitz and I. A. Stegun (eds.), *Handbook of Mathematical Functions* (Dover, New York, 1965).

[23] V. Azcoiti, G. di Carlo, A. Galante, A. F. Grillo, V. Laliena and C. E. Piedrafita, Zaragoza preprint DFTUZ/96/3 (1996) (`hep-lat/9601025`).

Table 1: Data for the average plaquette $P$. Listed are our old [2] and new data. In addition, on the $20^4$ lattice at $\beta = 0.20$ and $m = 0.005$ we obtain $P = 1.0504(5)$. We include (printed in *italics*) data from Refs. [13] ($8^4$) and [10] ($10^4$ and $16^4$).

| $\beta$ | $m$ | $L=4$ | $L=6$ | $L=8$ | $L=10$ | $L=12$ | $L=16$ |
|---|---|---|---|---|---|---|---|
| 0.160 | 0.02 | 1.3083(26) | 1.3538(12) | 1.3617(22) | | | |
| | 0.04 | 1.3266(39) | 1.3705(11) | 1.3739(11) | | | |
| | 0.05 | | | *1.378* | | | |
| | 0.09 | | | 1.4018(8) | | | |
| | 0.16 | | | 1.4280(8) | | | |
| 0.170 | 0.005 | | 1.2413(14) | 1.2528(10) | | 1.2529(7) | |
| | 0.01 | | 1.2461(12) | 1.2579(20) | | 1.2586(7) | |
| | 0.02 | 1.2237(26) | 1.2578(12) | 1.2669(12) | | 1.2669(6) | |
| | 0.03 | | | 1.2773(24) | | | |
| | 0.04 | 1.2362(19) | 1.2784(13) | 1.2837(13) | | 1.2832(6) | |
| | 0.05 | | | *1.287* | | | |
| | 0.09 | | | | | 1.3093(5) | |
| | 0.16 | | | | | 1.3376(4) | |
| 0.180 | 0.005 | | 1.1615(11) | 1.1684(14) | 1.1736(10) | 1.1729(5) | |
| | 0.01 | | 1.1669(11) | 1.1774(15) | | 1.1779(6) | |
| | 0.02 | 1.1566(16) | 1.1795(14) | 1.1873(20) | | 1.1881(6) | 1.1879(3) |
| | 0.03 | | | 1.1944(15) | | | |
| | 0.04 | 1.1692(15) | 1.1982(11) | 1.2031(9) | | 1.2022(6) | |
| | 0.05 | | | *1.208* | | | |
| | 0.09 | | | 1.2287(8) | | 1.2303(4) | |
| | 0.16 | | | 1.2583(5) | | 1.2577(4) | |
| 0.185 | 0.005 | | | 1.1323(9) | 1.1358(11) | 1.1363(17) | 1.1375(7) |
| | 0.01 | | 1.1327(11) | 1.1392(13) | | | |
| | 0.02 | | 1.1439(9) | 1.1482(12) | *1.1526(8)* | | |
| | 0.03 | | | 1.1588(13) | *1.1612(7)* | | |
| | 0.04 | 1.1360(16) | 1.1617(20) | 1.1654(12) | *1.1681(10)* | | |
| | 0.05 | | | 1.1738(11) | *1.1729(9)* | | |
| | 0.06 | | | | *1.1790(7)* | | |
| | 0.07 | | | | *1.1848(8)* | | |
| 0.190 | 0.005 | | | 1.1024(9) | 1.1037(12) | 1.1046(11) | 1.1057(5) |
| | 0.01 | | 1.1043(9) | 1.1099(8) | *1.1113(12)* | 1.1106(9) | 1.1108(3) |
| | 0.02 | | 1.1107(9) | 1.1166(11) | | 1.1194(5) | 1.1196(3) |
| | 0.03 | | | 1.1257(9) | *1.1282(9)* | | *1.1288(6)* |
| | 0.04 | 1.1049(14) | 1.1283(10) | 1.1331(13) | *1.1349(8)* | 1.1343(5) | |
| | 0.05 | | | *1.139* | 1.1424(9) | | |
| | 0.06 | | | | *1.1469(6)* | | |
| | 0.07 | | | | *1.1515(10)* | | |
| | 0.09 | | | 1.1589(6) | | 1.1610(3) | |
| | 0.16 | | | 1.1872(6) | | 1.1876(4) | |
| 0.195 | 0.005 | | | 1.0752(9) | 1.0766(8) | 1.0761(4) | 1.0769(3) |
| | 0.01 | | 1.0755(7) | 1.0773(11) | *1.0805(9)* | 1.0808(5) | 1.0817(2) |
| | 0.02 | | 1.0811(7) | 1.0879(13) | *1.0896(9)* | 1.0896(5) | 1.0902(3) |
| | 0.03 | | | 1.0946(10) | *1.0969(9)* | | *1.0992(5)* |
| | 0.04 | 1.0784(14) | 1.0954(7) | 1.1040(10) | *1.1036(10)* | 1.1026(6) | |
| | 0.05 | | | *1.109* | *1.1098(11)* | | |
| | 0.06 | | | | *1.1152(7)* | | |
| | 0.07 | | | | *1.1216(10)* | | |
| | 0.09 | | | | | 1.1296(5) | |

Table 1: Average plaquette data *(continued from previous page)*.

| $\beta$ | $m$ | $L=4$ | $L=6$ | $L=8$ | $L=10$ | $L=12$ | $L=16$ |
|---|---|---|---|---|---|---|---|
| 0.200 | 0.005 | | | 1.0506(9) | *1.0505(7)* | 1.0513(5) | 1.0506(2) |
| | 0.01 | | | 1.0539(7) | *1.0541(9)* | 1.0548(6) | 1.0548(2) |
| | 0.02 | | 1.0572(8) | 1.0592(6) | *1.0621(8)* | 1.0621(8) | 1.0617(2) |
| | 0.03 | | | 1.0677(10) | *1.0671(8)* | | *1.0710(5)* |
| | 0.04 | | 1.0680(7) | 1.0731(7) | 1.0766(7) | 1.0739(3) | |
| | 0.05 | | | *1.081* | 1.0810(8) | | |
| | 0.06 | | | | 1.0858(7) | | |
| | 0.07 | | | | *1.0913(6)* | | |
| | 0.09 | | | 1.1000(7) | | 1.0997(3) | |
| | 0.16 | | | 1.1244(5) | | 1.1256(3) | |
| 0.205 | 0.005 | | | | *1.0263(5)* | | |
| | 0.01 | | | 1.0291(6) | *1.0301(7)* | | |
| | 0.02 | | | | *1.0361(6)* | | |
| | 0.03 | | | 1.0401(8) | *1.0430(8)* | | *1.0442(4)* |
| | 0.04 | | | | *1.0485(7)* | | |
| | 0.05 | | | *1.054* | 1.0536(7) | | |
| | 0.06 | | | | 1.0585(6) | | |
| | 0.07 | | | | *1.0636(6)* | | |
| 0.210 | 0.005 | | | | *1.0060(5)* | | |
| | 0.01 | | | 1.0070(7) | *1.0073(6)* | 1.0068(4) | 1.0075(2) |
| | 0.02 | | 1.0084(6) | 1.0112(8) | 1.0135(6) | 1.0114(3) | 1.0127(2) |
| | 0.03 | | | 1.0160(7) | *1.0197(9)* | | *1.0181(4)* |
| | 0.04 | | 1.0178(6) | 1.0224(8) | 1.0247(10) | 1.0234(4) | |
| | 0.05 | | | *1.029* | 1.0287(5) | | |
| | 0.06 | | | | 1.0320(6) | | |
| | 0.07 | | | | *1.0374(7)* | | |
| | 0.09 | | | 1.0466(5) | | 1.0455(3) | |
| | 0.16 | | | | 1.0695(3) | | |
| 0.215 | 0.005 | | | | *0.9842(6)* | | |
| | 0.01 | | | 0.9861(9) | *0.9851(7)* | | |
| | 0.02 | | | | *0.9904(6)* | | |
| | 0.03 | | | 0.9946(8) | *0.9952(8)* | | *0.9950(3)* |
| | 0.04 | | | | *1.0003(7)* | | |
| | 0.05 | | | *1.003* | 1.0046(8) | | |
| | 0.06 | | | | 1.0087(8) | | |
| | 0.07 | | | | *1.0136(6)* | | |
| 0.220 | 0.005 | | | | *0.9638(6)* | | |
| | 0.01 | | | 0.9651(6) | *0.9661(5)* | 0.9657(4) | 0.9659(2) |
| | 0.02 | | 0.9664(6) | 0.9679(6) | *0.9679(4)* | 0.9692(3) | 0.9700(2) |
| | 0.03 | | | 0.9731(6) | *0.9735(10)* | | *0.9751(5)* |
| | 0.04 | | 0.9731(6) | 0.9771(8) | 0.9779(10) | 0.9779(3) | |
| | 0.05 | | | *0.980* | 0.9829(8) | | |
| | 0.06 | | | | 0.9869(6) | | |
| | 0.07 | | | | *0.9906(8)* | | |
| | 0.09 | | | 0.9972(4) | | 0.9978(2) | |
| | 0.16 | | | 1.0205(5) | | 1.0206(3) | |

Table 2: Data for the chiral condensate $\sigma$. Listed are our old [2] and new data as well as extrapolations to infinite $L$. In addition, on the $20^4$ lattice at $\beta = 0.20$ and $m = 0.005$ we obtain $\sigma = 0.0899(8)$. We include (printed in *italics*) data from Refs. [13] ($8^4$) and [10] ($10^4$ and $16^4$).

| $\beta$ | $m$ | $L=4$ | $L=6$ | $L=8$ | $L=10$ | $L=12$ | $L=16$ | $L=\infty$ |
|---|---|---|---|---|---|---|---|---|
| 0.160 | 0.02 | 0.1760(37) | 0.3611(13) | 0.3804(16) | | | | 0.3815(16) |
| | 0.04 | 0.2819(54) | 0.4012(9) | 0.4062(12) | | | | 0.4066(12) |
| | 0.05 | | | *0.418(3)* | | | | 0.4183(30) |
| | 0.09 | | | 0.4470(7) | | | | 0.4471(7) |
| | 0.16 | | | 0.4782(5) | | | | 0.4782(5) |
| 0.170 | 0.005 | | 0.1747(38) | 0.2699(17) | | 0.2921(9) | | 0.2904(8) |
| | 0.01 | | 0.2434(25) | 0.3047(27) | | 0.3109(13) | | 0.3109(11) |
| | 0.02 | 0.1289(40) | 0.3020(16) | 0.3325(13) | | 0.3329(9) | | 0.3336(7) |
| | 0.03 | | | 0.3547(17) | | | | 0.3559(17) |
| | 0.04 | 0.2185(26) | 0.3585(19) | 0.3689(8) | | 0.3698(7) | | 0.3697(5) |
| | 0.05 | | | *0.382(2)* | | | | 0.3825(20) |
| | 0.09 | | | | | 0.4194(3) | | 0.4194(3) |
| | 0.16 | | | | | 0.4577(3) | | 0.4577(3) |
| 0.180 | 0.005 | | 0.0977(30) | 0.1903(43) | 0.2177(26) | 0.2239(9) | | 0.2255(8) |
| | 0.01 | | 0.1684(30) | 0.2351(23) | | 0.2494(7) | | 0.2499(7) |
| | 0.02 | 0.0998(22) | 0.2473(32) | 0.2790(18) | | 0.2855(8) | 0.2843(4) | 0.2845(4) |
| | 0.03 | | | 0.3046(16) | | | | 0.3072(15) |
| | 0.04 | 0.1859(30) | 0.3143(20) | 0.3281(11) | | 0.3277(6) | | 0.3282(5) |
| | 0.05 | | | *0.344(2)* | | | | 0.3450(20) |
| | 0.09 | | | 0.3907(7) | | 0.3910(3) | | 0.3910(3) |
| | 0.16 | | | 0.4380(5) | | 0.4375(3) | | 0.4377(3) |
| 0.185 | 0.005 | | | 0.1383(41) | 0.1760(30) | 0.1843(25) | 0.1863(13) | 0.1872(10) |
| | 0.01 | | 0.1320(24) | 0.1963(34) | | | | 0.2176(29) |
| | 0.02 | | 0.2145(24) | 0.2501(20) | *0.2579(10)* | | | 0.2588(9) |
| | 0.03 | | | 0.2854(16) | *0.2887(10)* | | | 0.2891(8) |
| | 0.04 | 0.1687(31) | 0.2899(33) | 0.3069(13) | *0.3095(12)* | | | 0.3094(9) |
| | 0.05 | | | 0.3265(10) | *0.3261(9)* | | | 0.3267(6) |
| | 0.06 | | | | *0.3410(9)* | | | 0.3411(9) |
| | 0.07 | | | | *0.3554(8)* | | | 0.3555(8) |
| 0.190 | 0.005 | | | 0.1045(39) | 0.1348(29) | 0.1452(18) | 0.1512(9) | 0.1526(7) |
| | 0.01 | | 0.1121(26) | 0.1700(24) | 0.1820(21) | 0.1835(13) | 0.1885(5) | 0.1886(4) |
| | 0.02 | | 0.1841(30) | 0.2255(15) | 0.2282(16) | 0.2340(7) | 0.2334(4) | 0.2336(3) |
| | 0.03 | | | 0.2635(16) | *0.2650(12)* | | *0.2658(8)* | 0.2662(6) |
| | 0.04 | 0.1512(27) | 0.2656(22) | 0.2850(13) | *0.2893(14)* | 0.2892(6) | | 0.2891(5) |
| | 0.05 | | | *0.307(1)* | *0.3111(15)* | | | 0.3095(8) |
| | 0.06 | | | | *0.3257(8)* | | | 0.3258(8) |
| | 0.07 | | | | *0.3400(1)* | | | 0.3401(1) |
| | 0.09 | | | 0.3619(6) | | 0.3635(3) | | 0.3633(3) |
| | 0.16 | | | 0.4172(5) | | 0.4177(2) | | 0.4176(2) |
| 0.195 | 0.005 | | | 0.0834(31) | 0.1010(19) | 0.1114(11) | 0.1208(6) | 0.1226(5) |
| | 0.01 | | 0.0886(33) | 0.1305(18) | *0.1533(21)* | 0.1587(9) | 0.1600(3) | 0.1606(3) |
| | 0.02 | | 0.1555(15) | 0.1971(23) | *0.2065(14)* | 0.2102(5) | 0.2106(3) | 0.2107(2) |
| | 0.03 | | | 0.2399(24) | *0.2420(9)* | | *0.2450(6)* | 0.2445(5) |
| | 0.04 | 0.1389(27) | 0.2399(23) | 0.2677(13) | *0.2696(12)* | 0.2705(6) | | 0.2706(5) |
| | 0.05 | | | *0.289(1)* | *0.2905(13)* | | | 0.2911(8) |
| | 0.06 | | | | *0.3105(9)* | | | 0.3107(9) |
| | 0.07 | | | | *0.3271(8)* | | | 0.3272(8) |
| | 0.09 | | | | | 0.3508(3) | | 0.3508(3) |

Table 2: Chiral condensate data *(continued from previous page).*

| $\beta$ | $m$ | $L=4$ | $L=6$ | $L=8$ | $L=10$ | $L=12$ | $L=16$ | $L=\infty$ |
|---|---|---|---|---|---|---|---|---|
| 0.200 | 0.005 | | | 0.0580(16) | *0.0754(17)* | 0.0836(11) | 0.0902(6) | 0.0932(4) |
| | 0.01 | | | 0.1097(18) | *0.1244(18)* | 0.1322(10) | 0.1353(4) | 0.1364(4) |
| | 0.02 | | 0.1398(20) | 0.1694(10) | *0.1875(13)* | 0.1891(9) | 0.1876(2) | 0.1878(2) |
| | 0.03 | | | 0.2189(17) | *0.2214(13)* | | *0.2259(5)* | 0.2256(4) |
| | 0.04 | | 0.2226(11) | 0.2483(18) | 0.2523(9) | 0.2514(5) | | 0.2520(4) |
| | 0.05 | | | *0.273(3)* | 0.2754(8) | | | 0.2758(8) |
| | 0.06 | | | | *0.2938(6)* | | | 0.2941(6) |
| | 0.07 | | | | *0.3104(8)* | | | 0.3106(8) |
| | 0.09 | | | 0.3361(5) | | 0.3377(3) | | 0.3375(3) |
| | 0.16 | | | 0.3975(5) | | 0.3982(2) | | 0.3981(2) |
| 0.205 | 0.005 | | | | *0.0580(13)* | | | |
| | 0.01 | | 0.0892(13) | | *0.1059(10)* | | | 0.1190(11) |
| | 0.02 | | | | *0.1623(12)* | | | 0.1678(11) |
| | 0.03 | | | 0.1988(19) | *0.2058(13)* | | *0.2076(6)* | 0.2077(5) |
| | 0.04 | | | | *0.2325(13)* | | | 0.2337(13) |
| | 0.05 | | | *0.258(4)* | 0.2575(9) | | | 0.2583(9) |
| | 0.06 | | | | *0.2781(8)* | | | 0.2785(8) |
| | 0.07 | | | | *0.2969(8)* | | | 0.2971(8) |
| 0.210 | 0.005 | | | | *0.0491(12)* | | | |
| | 0.01 | | | 0.0788(15) | *0.0883(8)* | 0.0917(6) | 0.0974(4) | 0.0991(3) |
| | 0.02 | | 0.1083(18) | 0.1321(19) | *0.1466(10)* | 0.1467(5) | 0.1514(3) | 0.1511(2) |
| | 0.03 | | | 0.1789(18) | *0.1901(9)* | | *0.1899(6)* | 0.1908(5) |
| | 0.04 | | 0.1902(16) | 0.2126(18) | *0.2213(9)* | 0.2197(4) | | 0.2204(4) |
| | 0.05 | | | *0.242(4)* | 0.2434(11) | | | 0.2444(10) |
| | 0.06 | | | | *0.2640(9)* | | | 0.2645(9) |
| | 0.07 | | | | *0.2823(8)* | | | 0.2826(8) |
| | 0.09 | | | 0.3124(5) | | 0.3123(2) | | 0.3125(2) |
| | 0.16 | | | 0.3798(4) | | 0.3798(2) | | 0.3798(2) |
| 0.215 | 0.005 | | | | *0.0370(5)* | | | |
| | 0.01 | | 0.0648(10) | | *0.0729(7)* | | | |
| | 0.02 | | | | *0.1292(8)* | | | 0.1365(8) |
| | 0.03 | | | 0.1624(14) | *0.1737(10)* | | *0.1746(5)* | 0.1751(4) |
| | 0.04 | | | | *0.2054(11)* | | | 0.2072(11) |
| | 0.05 | | | *0.226(3)* | 0.2303(12) | | | 0.2313(11) |
| | 0.06 | | | | *0.2519(9)* | | | 0.2525(9) |
| | 0.07 | | | | *0.2695(8)* | | | 0.2699(8) |
| 0.220 | 0.005 | | | | *0.0325(4)* | | | |
| | 0.01 | | | 0.0570(5) | 0.0658(6) | 0.0692(8) | 0.0722(2) | 0.0747(2) |
| | 0.02 | | 0.0893(10) | 0.1012(34) | *0.1144(7)* | 0.1213(4) | 0.1223(2) | 0.1233(2) |
| | 0.03 | | | 0.1527(10) | *0.1588(16)* | | *0.1625(5)* | 0.1631(4) |
| | 0.04 | | 0.1616(15) | 0.1808(10) | *0.1921(15)* | 0.1917(4) | | 0.1920(4) |
| | 0.05 | | | *0.210(3)* | 0.2173(10) | | | 0.2183(9) |
| | 0.06 | | | | *0.2387(9)* | | | 0.2395(9) |
| | 0.07 | | | | *0.2572(8)* | | | 0.2577(8) |
| | 0.09 | | | 0.2887(5) | | 0.2898(2) | | 0.2899(2) |
| | 0.16 | | | 0.3618(5) | | 0.3624(2) | | 0.3624(2) |

Table 3: Data for the susceptibility ratio $R$. All of these data are new. In addition, on the $20^4$ lattice at $\beta = 0.20$ and $m = 0.005$ we obtain $R = 0.642(33)$.

| $\beta$ | $m$ | $L=4$ | $L=6$ | $L=8$ | $L=10$ | $L=12$ | $L=16$ | $L=\infty$ |
|---|---|---|---|---|---|---|---|---|
| 0.160 | 0.02 | 0.802(8) | 0.153(3) | | | | | |
| | 0.04 | 0.554(12) | 0.137(4) | | | | | |
| 0.170 | 0.005 | | 0.652(12) | 0.169(11) | | 0.077(3) | | 0.077(3) |
| | 0.01 | | 0.386(12) | 0.130(6) | | 0.093(3) | | 0.095(3) |
| | 0.02 | 0.905(10) | 0.253(5) | 0.143(2) | | 0.130(4) | | 0.131(2) |
| | 0.03 | | | 0.142(7) | | | | 0.136(7) |
| | 0.04 | 0.718(6) | 0.205(6) | 0.149(3) | | 0.164(11) | | 0.146(3) |
| | 0.09 | | | 0.161(5) | | | | 0.161(5) |
| 0.180 | 0.005 | | 0.869(8) | 0.329(33) | 0.177(11) | 0.147(6) | | 0.130(5) |
| | 0.01 | | 0.635(12) | 0.260(9) | | 0.158(8) | | 0.158(6) |
| | 0.02 | 0.934(5) | 0.415(10) | | | 0.194(6) | 0.195(7) | 0.194(5) |
| | 0.03 | | | 0.254(5) | | | | 0.236(5) |
| | 0.04 | 0.803(11) | 0.303(8) | | | 0.222(9) | | 0.222(9) |
| | 0.09 | | | 0.214(6) | | | | 0.214(6) |
| 0.185 | 0.005 | | | 0.549(55) | 0.295(13) | 0.170(16) | 0.186(18) | 0.168(8) |
| | 0.01 | | 0.769(10) | 0.395(28) | | | | 0.221(21) |
| | 0.02 | | 0.518(10) | 0.295(15) | | | | 0.240(13) |
| | 0.03 | | | 0.285(14) | | | | 0.260(13) |
| | 0.04 | 0.838(7) | 0.362(13) | 0.279(10) | | | | 0.264(10) |
| | 0.05 | | | 0.264(5) | | | | 0.255(5) |
| 0.190 | 0.005 | | | 0.704(46) | 0.487(35) | 0.377(20) | 0.314(15) | 0.290(10) |
| | 0.01 | | 0.834(11) | 0.494(18) | | 0.336(21) | 0.308(12) | 0.293(8) |
| | 0.02 | | 0.631(13) | | | 0.309(7) | 0.316(15) | 0.306(6) |
| | 0.03 | | | 0.336(11) | | | | 0.299(10) |
| | 0.04 | 0.873(4) | 0.438(11) | | | 0.298(5) | | 0.297(5) |
| | 0.09 | | | | | 0.262(3) | | 0.262(3) |
| 0.195 | 0.005 | | | 0.778(36) | 0.651(19) | 0.554(16) | 0.410(14) | 0.394(9) |
| | 0.01 | | 0.876(31) | 0.672(21) | | 0.435(18) | 0.398(6) | 0.393(5) |
| | 0.02 | | 0.738(7) | 0.492(30) | | 0.377(8) | 0.366(3) | 0.365(3) |
| | 0.03 | | | 0.412(22) | | | | 0.356(20) |
| | 0.04 | 0.893(4) | 0.527(11) | 0.373(12) | | 0.346(4) | | 0.344(4) |
| | 0.09 | | | | | 0.292(7) | | 0.292(7) |
| 0.200 | 0.005 | | | 0.903(27) | | 0.763(16) | 0.636(21) | 0.584(13) |
| | 0.01 | | | 0.726(45) | | 0.555(20) | 0.505(11) | 0.482(9) |
| | 0.02 | | 0.789(13) | | | 0.462(12) | 0.446(4) | 0.444(4) |
| | 0.03 | | | 0.487(14) | | | | 0.410(13) |
| | 0.04 | | 0.582(8) | | | 0.411(5) | | 0.409(5) |
| | 0.09 | | | | | 0.321(6) | | 0.321(6) |
| 0.205 | 0.01 | | | 0.863(14) | | | | |
| | 0.03 | | | 0.565(22) | | | | 0.466(20) |
| 0.210 | 0.01 | | | 0.881(16) | | 0.780(9) | 0.691(13) | 0.659(8) |
| | 0.02 | | 0.881(6) | 0.747(34) | | 0.653(8) | 0.593(13) | 0.599(6) |
| | 0.03 | | | 0.641(19) | | | | 0.522(17) |
| | 0.04 | | 0.690(8) | 0.571(14) | | 0.499(12) | | 0.495(9) |
| | 0.09 | | | | | 0.386(6) | | 0.386(6) |
| 0.215 | 0.01 | | | 0.937(17) | | | | |
| | 0.03 | | | 0.701(26) | | | | 0.570(23) |
| 0.220 | 0.01 | | | 0.945(3) | | 0.887(13) | 0.836(5) | 0.774(6) |
| | 0.02 | | 0.927(2) | | | 0.742(7) | 0.709(4) | 0.685(4) |
| | 0.03 | | | 0.728(14) | | | | 0.594(13) |
| | 0.04 | | 0.792(5) | | | 0.583(11) | | 0.571(11) |
| | 0.09 | | | | | 0.435(7) | | 0.434(7) |

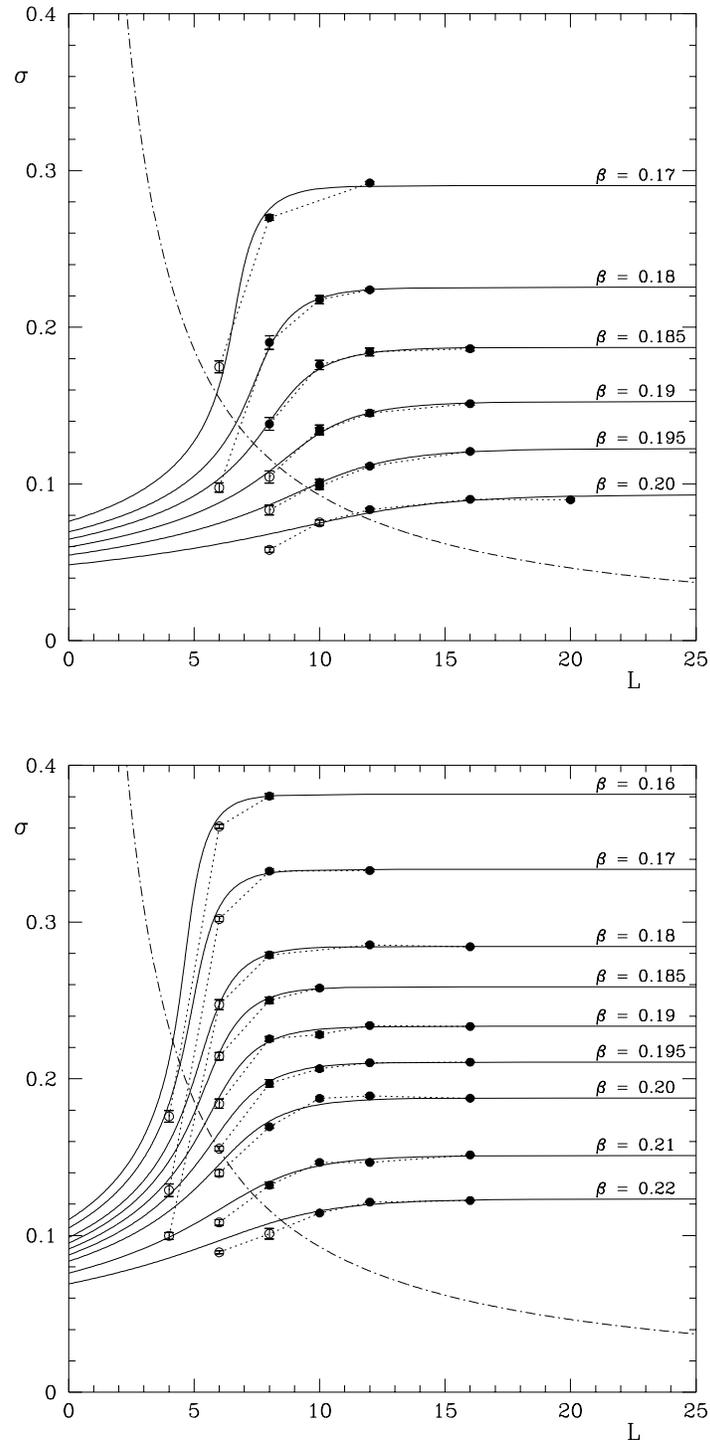

Figure 1: Finite size scaling behaviour of $\sigma$ at $m = 0.005$ (top) and $m = 0.02$ (bottom). Solid (open) symbols denote whether a point is (not) included in the fit. The solid lines show the fit to Eq. (25). The dashed-dotted line represents $3.23\,\sigma L = 3$.

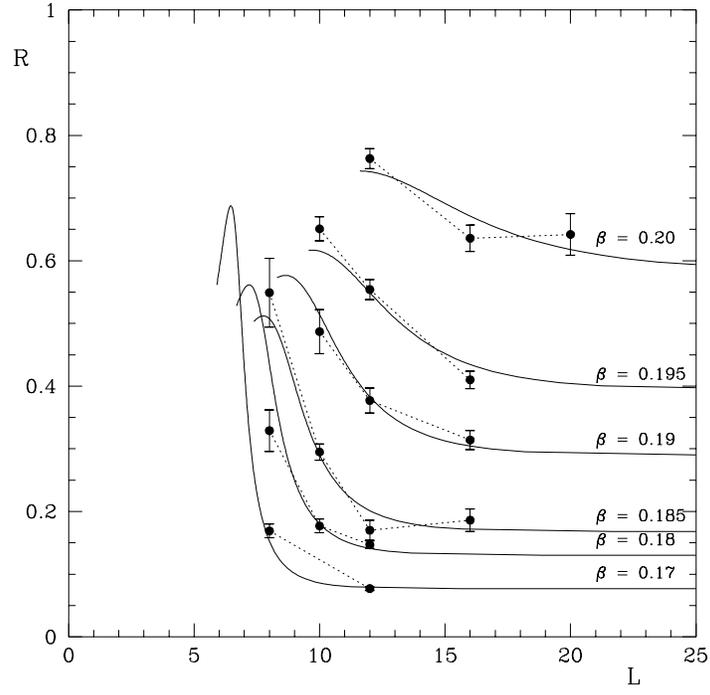

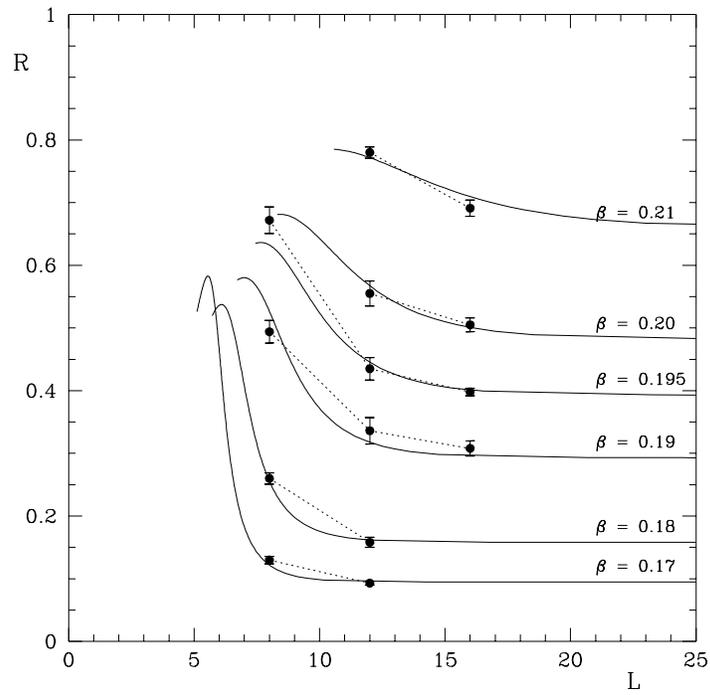

Figure 2: Finite size scaling behaviour of $R$ at $m = 0.005$ (top) and $m = 0.01$ (bottom). Shown are data that are included in the fit. The lines are the fit to Eq. (27). The lines are plotted in the region $3.23\,\sigma L > 3$.

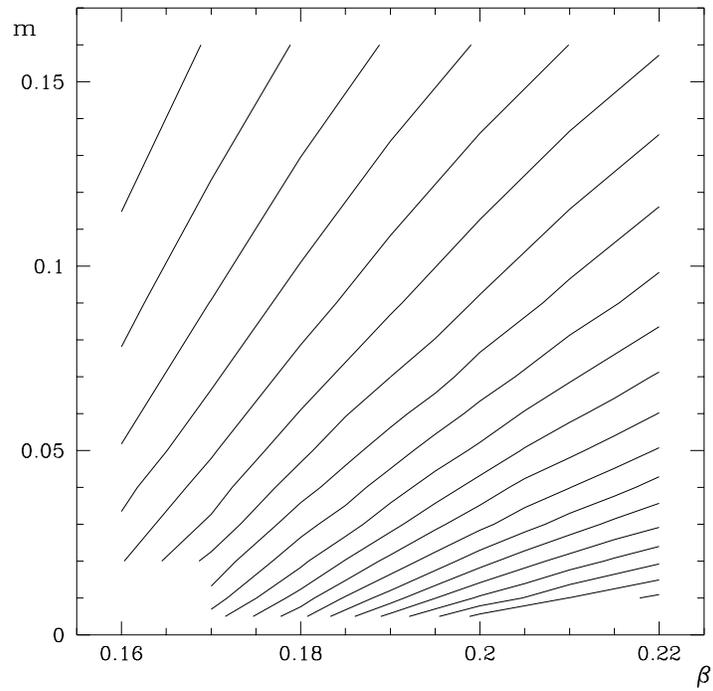

Figure 3: Contours of constant values of $\sigma$ in the $(\beta, m)$-plane. The values are $\sigma = 0.08$ to $\sigma = 0.46$ in steps of $0.02$, from bottom right to top left.

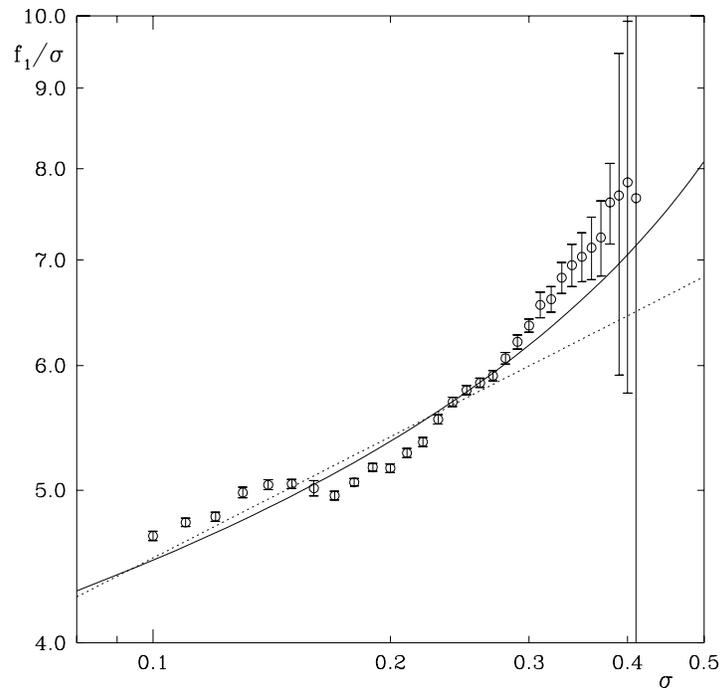

Figure 4: Numerical results for $f_1$ (defined in Eq. (29)). The lines represent fits discussed in section 5. The solid line represents fit 1, the dotted line represents fit 3.

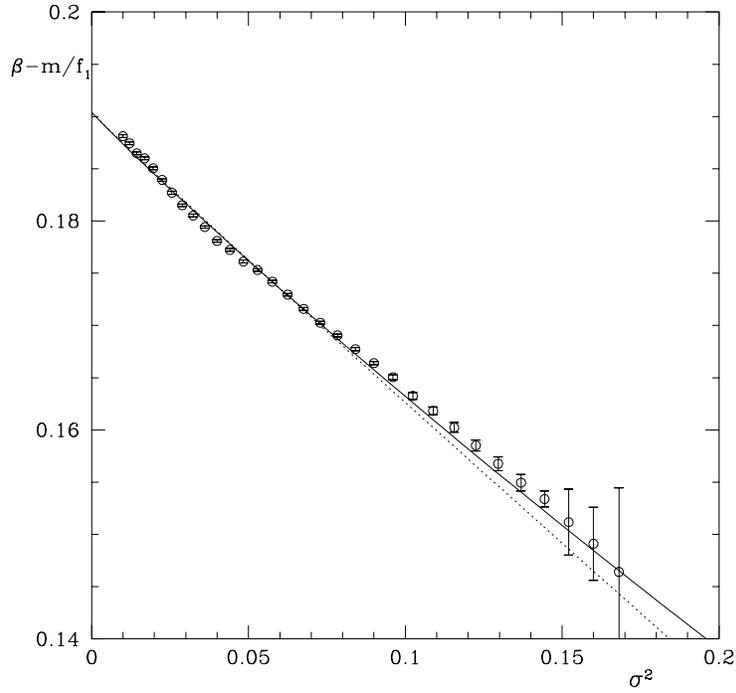

Figure 5: Numerical results for the l.h.s. of Eq. (31), allowing one to read off $\beta_c$ from the intercept. The lines represent fits discussed in section 5. The solid line represents fit 1, the dotted line represents fit 3.

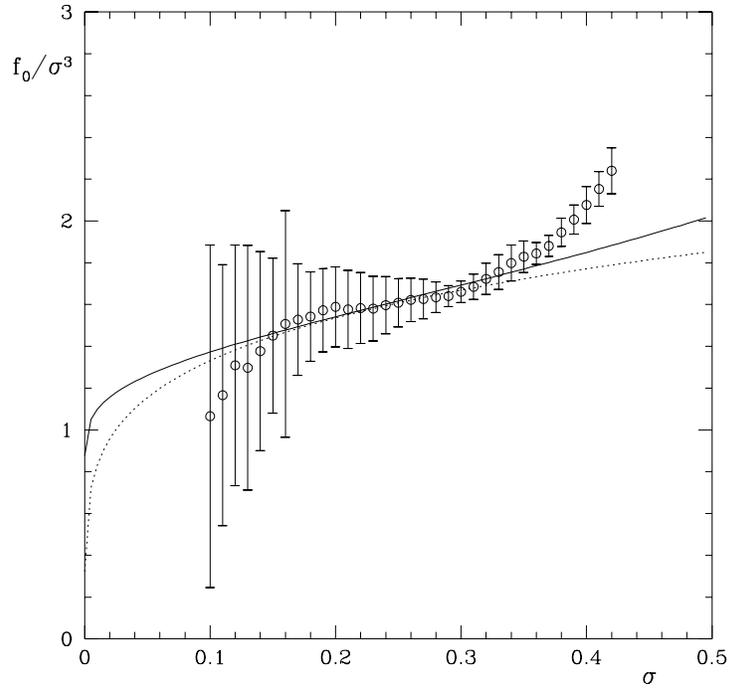

Figure 6: Numerical results for $f_0$ (defined in Eq. (29)). The lines represent fits discussed in section 5. The solid line represents fit 1, the dotted line represents fit 3.

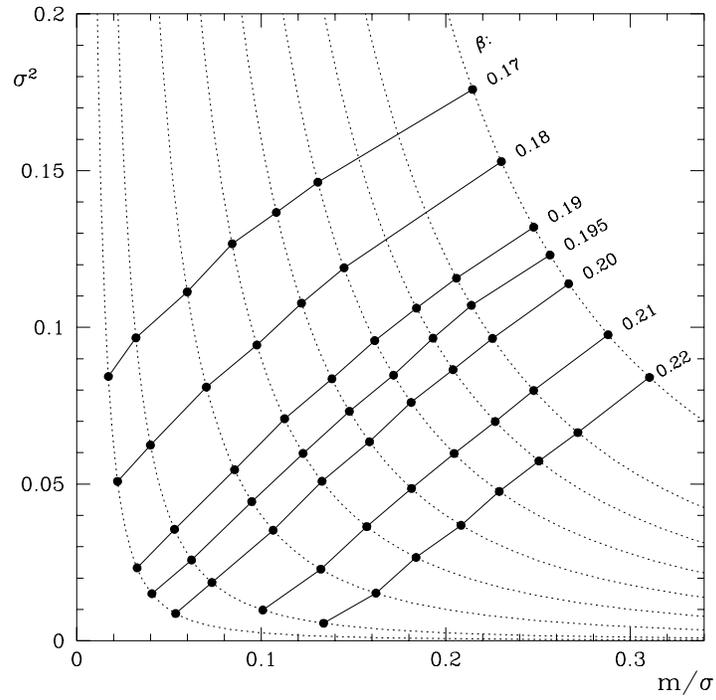

Figure 7: Fisher plot of extrapolated $\sigma$ data. The solid lines connect data belonging to the same $\beta$. The dotted lines are lines of constant $m$. The values are $m = 0.005, 0.01, 0.02, 0.03, 0.04, 0.05, 0.06, 0.07$ and $0.09$, from left to right.

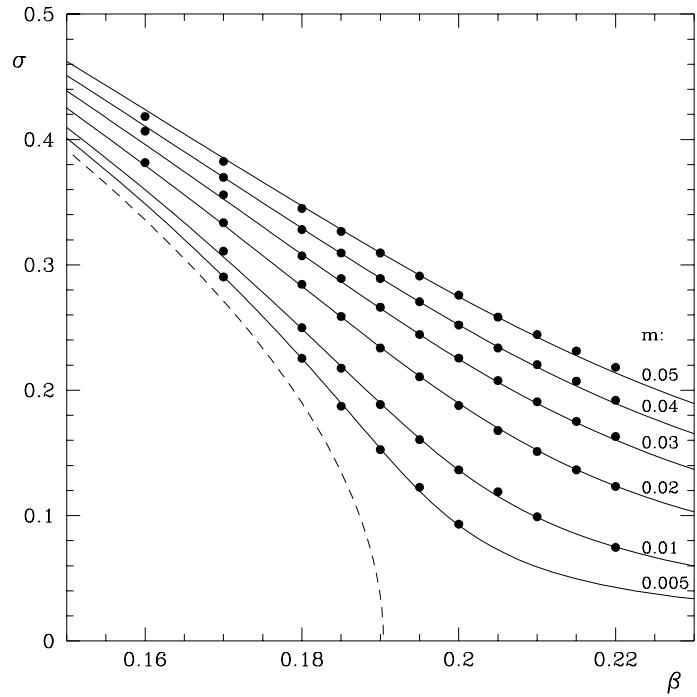

Figure 8: Fit of the logarithmically improved equation of state (34) to $\sigma$ data. The dashed line is the extrapolation to $m = 0$.

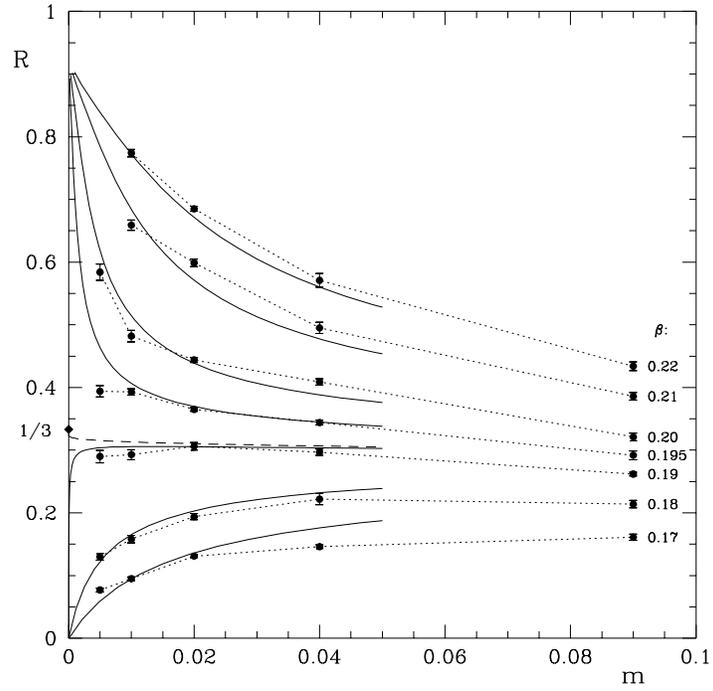

Figure 9: Plot of $R$ data. The dotted lines shall guide the eye. The solid lines represent the fit to the logarithmically improved equation of state (34) (fit 1 from section 5). The dashed line corresponds to $\beta_c$ from that fit.

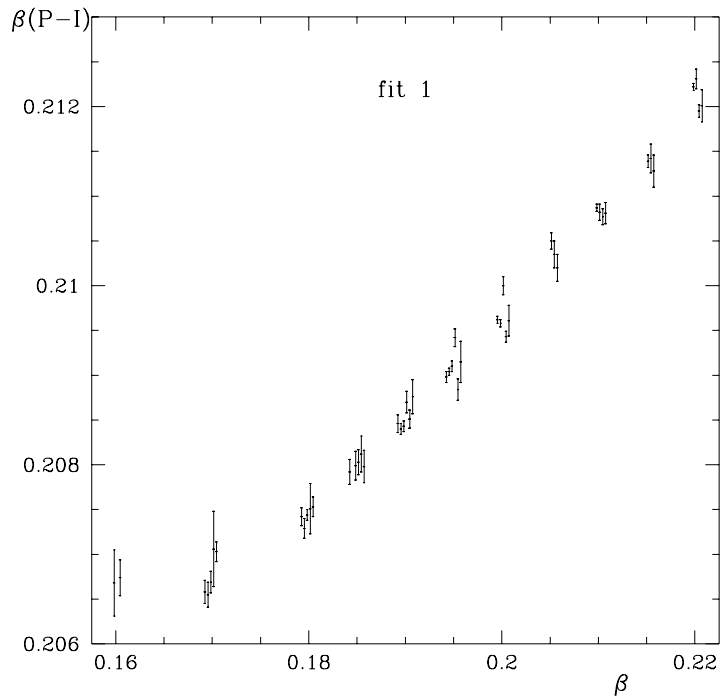

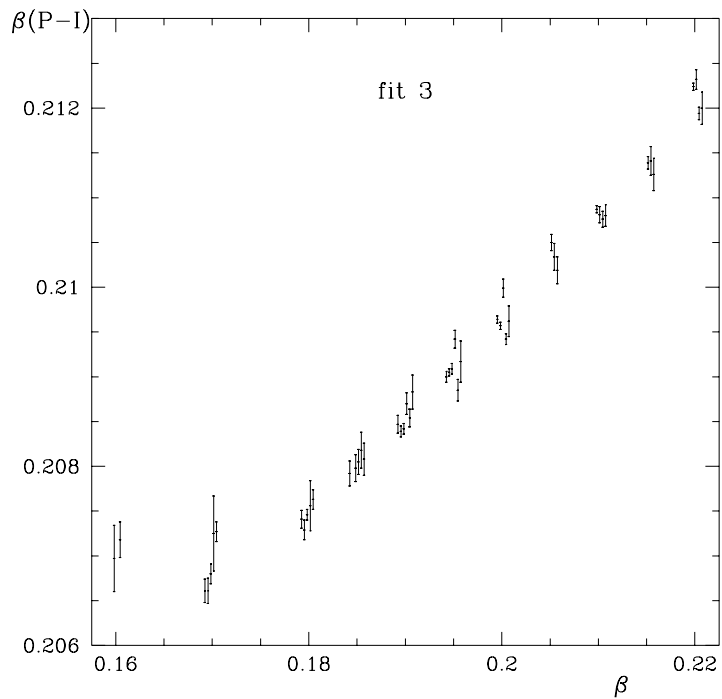

Figure 10: Test of the mass independence of $\beta(P(\sigma,\beta) - I(\sigma))$ for the best two fits of section 5. The mass $m$ takes values in the range $0.005 \leq m \leq 0.05$. Average plaquette data from the largest available lattices are used. Results for a given $\beta$ are plotted beside one another, with $m$ growing from left to right.

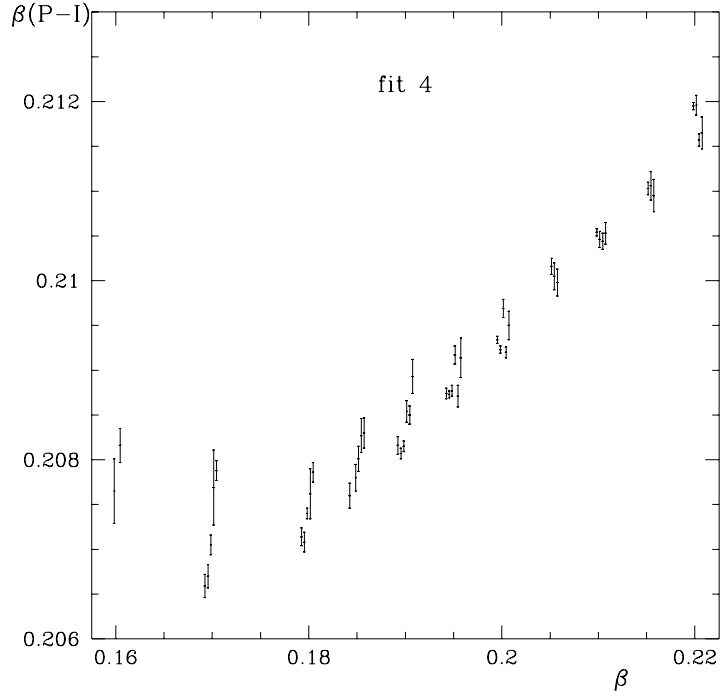

Figure 11: Test of the mass independence of $\beta(P(\sigma,\beta)-I(\sigma))$ for fit 4 (power-law equation of state with $b \equiv 1$ imposed). In this case the $m$ dependence is considerably greater than in Fig. 10.

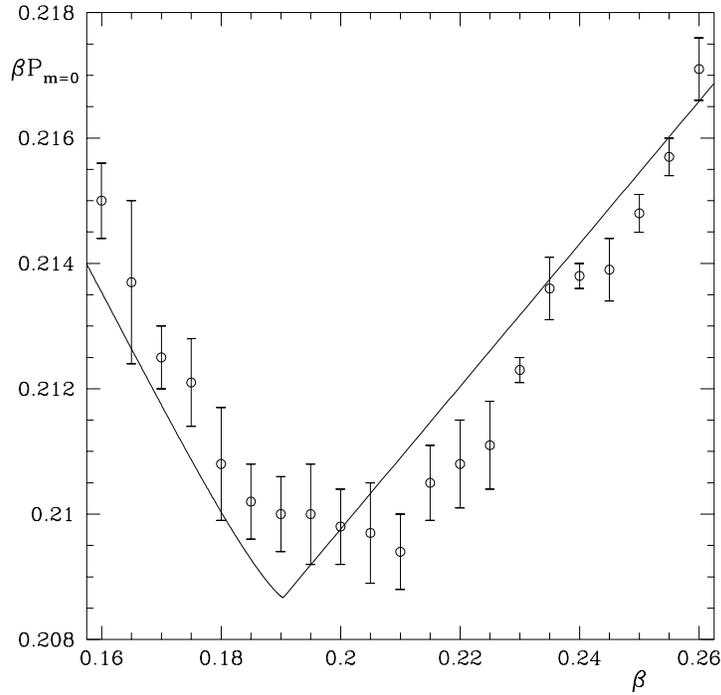

Figure 12: Comparison of the average plaquette data from the Zaragoza group on an $8^4$ lattice at $m = 0$ [13] with infinite volume values expected by the Maxwell relation (40). To draw the line we used the parameters of fit 1 and took $C(\beta) = 0.1138 + 0.1870/\beta$.